\documentclass[10pt,aps,preprint]{revtex4-1}

\usepackage{amssymb}
\usepackage{amsmath}
\usepackage{graphics}
\usepackage{latexsym}
\usepackage{array}
\usepackage[activeacute,english]{babel}
\usepackage[latin1]{inputenc}
\usepackage[dvips]{graphicx}
\usepackage{color}
\usepackage{epstopdf}
\usepackage{amssymb}
\usepackage{slashed}
\usepackage{mathrsfs}
\usepackage{amsmath}
\usepackage{wasysym}

\newcommand{\bn}{\begin{eqnarray}}
\newcommand{\en}{\end{eqnarray}}
\newcommand{\be}{\begin{equation}}
\newcommand{\ee}{\end{equation}}

\newcommand{\bc}{\begin{center}}
\newcommand{\ec}{\end{center}}

\begin{document}

\title{\Large \bf Graphene physics via the Dirac oscillator in (2+1) dimensions}

\author{C. Quimbay\footnote{Associate researcher of Centro
Internacional de F\'{\i}sica, Bogot\'{a} D.C., Colombia.}}
\email{cjquimbayh@unal.edu.co}

\affiliation{Departamento de F\'{\i}sica, Universidad Nacional de Colombia.\\
Ciudad Universitaria, Bogot\'{a} D.C., Colombia.}

\author{P. Strange}
\email{P.Strange@kent.ac.uk}

\affiliation{School of Physical Science, University of Kent.\\
Canterbury, United Kingdom}

\date{\today}

\begin{abstract}
We show how the two-dimensional Dirac oscillator model can
describe some properties of electrons in graphene. This model
explains the origin of the left-handed chirality observed for
charge carriers in monolayer and bilayer graphene. The
relativistic dispersion relation observed for monolayer graphene
is obtained directly from the energy spectrum, while the parabolic
dispersion relation observed for the case of bilayer graphene is
obtained in the non-relativistic limit. Additionally, if an
external magnetic field is applied, the unusual Landau-level
spectrum for monolayer graphene is obtained, but for bilayer
graphene the model predicts the existence of a magnetic
field-dependent gap. Finally, this model also leads to
the existence of a chiral phase transition.\\

\vspace{0.4cm} \noindent {\it{Keywords:}} {Dirac oscillator in
(2+1) dimensions, electronic properties of graphene, monolayer and
bilayer graphene, dispersion relations, Landau levels.}

\vspace{0.4cm} \noindent {\it{PACS Codes:}}  03.65.Pm, 72.80.Vp,
71.70.Di.

\end{abstract}

\maketitle

The three-dimensional system defined by the Dirac equation in the
presence of a linear vector potential of the form:
$im\omega\beta\vec{\alpha} \cdot\vec{r}$, where $m$ is the fermion
mass, $\omega$ an oscillator frequency, $\vec{r}$ the vector
distance of the fermion from the origin of the potential, and
$\vec{\alpha}$ and $\beta$ are the four Dirac matrices, has been
called the Dirac oscillator \cite{MS89} because it behaves in the
non-relativistic limit as an harmonic oscillator with a strong
spin-orbit coupling \cite{ST98}. Different physics and
mathematical aspects of this system have been widely studied in
one, two and three dimensions \cite{MM89}-\cite{BA2013}.
Additionally, this system has been used in quantum optics to study
properties of the (Anti)-Jaynes-Cummings model
\cite{AB07-1}-\cite{BO2013}. After more than 20 years of
theoretical activity focussed in the characterization of the Dirac
oscillator, just recently a first experimental realization of this
system was developed \cite{FV2013}. However, until now the Dirac
oscillator does not describe a known physical system
\cite{FV2013}.

On the other hand, the study of the electronic properties for
monolayer and bilayer graphene has been of huge interest in recent
years (see, for instance,\cite{CA2009,DAS2011,MC2013} and
references therein). It has been observed that electric charge
carriers of monolayer graphene behave like massless relativistic
fermions that obey a linear dispersion relation
\cite{CA2009,DAS2011}, while the electric charge carriers for the
case of bilayer graphene behave like massive non-relativistic
fermions that obey a quadratic dispersion relation
\cite{DAS2011,MC2013}. For both systems, the explanation of the
origin of these dispersion relations comes from the traditional
tight-binding approach. However, the tight-binding Hamiltonians
for electrons considering that electrons can hope to both nearest-
and next-nearest-neighbor are different for both systems
\cite{CA2009,DAS2011,MC2013}, thus this approach leads to
different explanations of the origin of the distinct dispersion
relations are of the two systems. In addition to the different
behaviour of the energy spectra, it is difficult to understand how
the massive electric charge carriers have a defined left-handed
chirality for the case of bilayer graphene.

The main goal of this letter is to show for the first time that
the Dirac oscillator can describe a naturally occurring physical
system. Specifically, this work shows that the (2+1)-dimensional
Dirac oscillator can be used to describe the dynamics of the
charge carriers in graphene, and hence its electronic properties.
To do this, we propose a model in which the electromagnetic
interaction of the medium over charge carriers of graphene is
described by means of a linear vector potential in the Dirac
equation. Solution of the model shows that the same formalism can
describe the linear relativistic dispersion relation for massless
charge carriers of monolayer graphene and the parabolic dispersion
relation for massive charge carriers of bilayer graphene. The
linear term that appears in the Dirac oscillator model leads
directly to the left-handed chirality observed in the charge
carriers for both monolayer and bilayer graphene. We also show
that if an external and uniform magnetic field is present in the
system, the model describes the unusual Landau levels observed for
the case of monolayer graphene, while for bilayer graphene it
predicts the existence of a magnetic field-dependent gap. Finally,
we show that this model predicts that changing the strength of the
magnetic field must lead to the existence of a chiral quantum
phase transition for the system.

We propose a Dirac oscillator in (2+1) dimensions for which the
linear potential has the form: $ieB_{I}\sigma_z
\vec{\sigma}\cdot\vec{r}/4$, where $\vec{r}=(x,y)$,
$\alpha_x=\sigma_x$, $\alpha_y=\sigma_y$, $\beta=\sigma_z$, $e$ is
the electrical charge of electron and $B_{I}$ is an internal
uniform effective magnetic field which is perpendicular to the
plane $(\vec B_I = B_I \hat e_z)$ and is assumed to originate in
an effective form from the motion of the charge carriers relative
to the planar hexagonal arrangement of carbon atoms. The model is
based on the premise that the dynamics of charge carriers in
graphene can be effectively described via this two-dimensional
Dirac oscillator
\begin{align}\label{eq:de1}
i\hbar \frac{\partial}{\partial t} \mid \psi > =\left[v_f
\sum_{i=1}^{2}\sigma_j\left(p_j - i\frac{eB_{I}}{4}\sigma_z r_j
\right) + \sigma_z m v_f^2 \right]\mid \psi >,
\end{align}
where $v_f$ is the Fermi speed of the charge carriers, $m$ is the
effective mass, $p_j$ are the components of the linear momentum,
$r_j$ are the spatial coordinates in the $(x,y)$ plane with
respect to the origin of the potential, $\sigma_j$ are the
non-diagonal Pauli matrices, $\sigma_z$ is the diagonal Pauli
matrix. The assumption about the origin of $B_I$ can be understood
in analogous way to the internal magnetic field that determines
the spin-orbit coupling of the hydrogen atom, which arises from
the relative motion of the electron relative to the proton. It is
well known that for the case of monolayer graphene in a
perpendicular magnetic field \cite{CA2009}, the Hamiltonian
$H_{ml} = v_f[\tau \sigma_x(p_x -ieB_I\sigma_z x/4)+ \sigma_y(p_y
-ieB_I\sigma_z y/4)]$ describes the single-particle states around
one of the two equivalent corners of the first Brillouin zone.
$\tau =\pm 1$ determines which corner of the Brillouin zone is
considered \cite{SC2008}. This fact means that we are considering
in Eq. (\ref{eq:de1}) the corner defined by $\tau = 1$.

For the case of monolayer graphene, the charge carriers are
considered massless $(m=0)$ and the mass term in Eq.
(\ref{eq:de1}) does not exist. The spinor $\mid \psi >$ is written
as $\mid \psi > = (\mid \psi_1>,\mid \psi_2 > )^T$. Substituting
this spinor in Eq. (\ref{eq:de1}), we obtain the following two
coupled equations
\begin{align}
(E-m v_f^2) \mid \psi_1 > &= v_f \left[(p_x + i\frac{\hbar
\omega^2}{v_f^2}x)- i(p_y + i\frac{\hbar \omega^2}{v_f^2}y)\right]
\mid \psi_2>, \label{eq:de11}\\
(E+m v_f^2) \mid \psi_2 > &= v_f \left[(p_x - i\frac{\hbar
\omega^2}{v_f^2}x)+ i(p_y - i\frac{\hbar \omega^2}{v_f^2}y)\right]
\mid \psi_1>,\label{eq:de12}
\end{align}
where $\omega^2=v_f^2/(4l_B^2)$, with $l_B^2=\hbar/(eB_I)$. The
quantity $l_B$ represents a new length scale in the problem
associated to the quantity $eB_I$ that appears in the linear
potential. In order to find the solutions of Eqs. (\ref{eq:de11})
and (\ref{eq:de12}), we introduce the right-handed chiral
annihilation and creation operators given by $a_r =\frac{1}{\sqrt
2}(a_x-ia_y)$, $a_r^\dag =\frac{1}{\sqrt 2}(a_x^\dag+ia_y^\dag)$
and the left-handed chiral annihilation and creation operators
given by $a_l =\frac{1}{\sqrt 2}(a_x+ia_y)$, $a_l^\dag
=\frac{1}{\sqrt 2}(a_x^\dag-ia_y^\dag)$, where $a_x$, $a_y$,
$a_x^\dag$ and $a_y^\dag$ are the usual annihilation and creation
operators of the harmonic oscillator defined respectively as $a_j
=\frac{1}{\sqrt 2}\left(\frac{1}{\Delta} r_j +
i\frac{\Delta}{\hbar} p_j\right)$ and $a_j^\dag =\frac{1}{\sqrt
2}\left(\frac{1}{\Delta} r_j - i\frac{\Delta}{\hbar} p_j\right)$,
with $\Delta = v_f/\omega$ representing the ground-state
oscillator width. Because the orbital angular momentum $L_z$ is
written in terms of the chiral annihilation and creation operators
as $L_z=\hbar(a_r^\dag a_r - a_l^\dag a_l)$, then $a_r^\dag$ and
$a_l^\dag$ are interpreted as the operators that create a right or
left quantum of angular momentum, respectively \cite{AB07-1}.
After expressing $r_j$ and $p_j$ in terms of operators $a_j$ and
$a_j^\dag$, the Eqs. (\ref{eq:de11}) and (\ref{eq:de12}) lead to
\begin{align}
\mid \psi_1 > &= i \frac{2\hbar\omega}{E-m v_f^2} a_l^\dag
\mid \psi_2>, \label{eq:de13}\\
\mid \psi_2 > &= - i\frac{2\hbar\omega}{E+m v_f^2} a_l \mid
\psi_1>,\label{eq:de14}
\end{align}
where only the left-handed chiral operators are present
\cite{AB07-1}. If we substitute Eqs. (\ref{eq:de13}) and
(\ref{eq:de14}) in (\ref{eq:de1}), the Hamiltonian describing the
dynamics of charge carriers in graphene can be written as
\begin{align}
H_1= \hbar ( g_l \sigma^{+} a_l^\dag + g_l^* \sigma^{-} a_l )
+mv_f^2\sigma_z, \label{eq:Doham1}
\end{align}
with $\sigma^+$ and $\sigma^-$ representing the spin raising and
lowering operators and $g_l = i2\omega=iv_f\sqrt{eB_I/\hbar}$. The
two-dimensional Hamiltonian $H_1$ describes a massive charge
carrier of left-handed chirality.

From Eqs. (\ref{eq:de13}) and (\ref{eq:de14}), it is possible to
write the associated Klein-Gordon equations and then to obtain the
energy spectrum for charge carriers of left-handed chirality
\begin{align}
E_{n_l}= \pm \sqrt{\hbar^2k^2v_f^2\left(n_l + \frac{1}{2} \mp
\frac{1}{2}\right) + m^2v_f^4}, \label{eq:enes1}
\end{align}
with $n_l=0,1,2,\ldots$ and $k^2=e B_I/\hbar$. For the case in
which the effective mass of charge carriers vanishes ($m=0$), the
energy spectrum (\ref{eq:enes1}) is written as
\begin{align}
E_{n_l}= \pm \hbar kv_f \sqrt{n_l + \frac{1}{2} \mp \frac{1}{2}},
\label{eq:enes2}
\end{align}
which corresponds to the relativistic dispersion relation
depending linearly on the momentum $\hbar k$ and which is
consistent with the well known result for charge carriers in
monolayer graphene \cite{CA2009,DAS2011}. On the other hand, for
the case where the effective mass of the charge carriers is
non-zero ($m\neq0$), the non-relativistic limit of
(\ref{eq:enes1}) yields
\begin{align}
E^{nr}_{n_l}= \pm \sqrt{\hbar^2 k^2v_f^2\left(n_l + \frac{1}{2}
\mp \frac{1}{2}\right) + m^2v_f^4} - mv_f^2. \label{eq:enes3}
\end{align}
For this limit, it is valid to write $mv_f\gg \hbar k$ and then
the energy spectrum for charge carriers can be written as
\begin{align}
E^{+}_{n_l}= \frac{\hbar^2k^2}{2m}\left(n_l + \frac{1}{2} \mp
\frac{1}{2}\right), \label{eq:enes4}
\end{align}
which corresponds to the non-relativistic dispersion relation
depending parabolically on the momentum, which is observed for
charge carriers in bilayer graphene \cite{NO2006}. We note that,
within the same theoretical formalism, we have been able to obtain
the very well known dispersion relations of charge carriers in
monolayer and bilayer graphene, that are usually obtained starting
from the standard nearest-neighbour approximation using two
different tight-binding Hamiltonians \cite{CA2009,DAS2011,MC2013}.
Because the linear term of the Dirac oscillator leads directly to
a description of a fermion with an intrinsic left-handed
chirality, which is only present in the two-dimensional Dirac
oscillator \cite{AB07-1}, the charge carriers of graphene describe
by Eq. (\ref{eq:de1}) are left-handed. In this sense, the Dirac
oscillator described by equation (\ref{eq:de1}) can explain the
left-handed chirality observe for massless and massive charge
carriers in graphene.

Next we consider the effect of an external uniform magnetic field
on the system. The dynamics of the charge carriers of graphene are
then described by
\begin{align}\label{eq:de2}
i\hbar \frac{\partial}{\partial t} \mid \psi > =\left[v_f
\sum_{i=1}^{2}\sigma_j\left(p_j - i\frac{eB_{I}}{4}\sigma_z r_j
-eA_j \right) + \sigma_z m v_f^2 \right]\mid \psi >,
\end{align}
where we are considering the corner of the first Brillouin zone
defined by $\tau = 1$ \cite{SC2008}. We observe that for the case
of a vanishing magnetic field ($A_j=0$) in Eq. (\ref{eq:de2}),
then the Eq. (\ref{eq:de1}) is recovered. In order to understand
the physics described by Eq. (\ref{eq:de2}), we assume initially
that the linear potential in vanishes. This fact means we are
considering an charge carriers with mass in presence of an
external magnetic field. i.e.
\begin{align}\label{eq:de3}
i\hbar \frac{\partial}{\partial t} \mid \psi > =\left[v_f
\sum_{i=1}^{2}\sigma_j\left(p_j -eA_j \right) + \sigma_z m v_f^2
\right]\mid \psi >.
\end{align}
Substituting the spinor $\mid \psi > = (\mid \psi_1 >,\mid \psi_2
> )^T$ in Eq. (\ref{eq:de3}), yields following two coupling
equations:
\begin{align}
(E-m v_f^2) \mid \psi_1 > &= v_f [(p_x -eA_x)- i(p_y - eA_y)]
\mid \psi_2>, \label{eq:de31}\\
(E+m v_f^2) \mid \psi_2 > &= v_f [(p_x -eA_x)+ i(p_y - eA_y)] \mid
\psi_1>,\label{eq:de32}
\end{align}
The external magnetic field, which is perpendicular to the plane,
is written as $\vec B = - B \hat e_z$. The vector potential $\vec
A$ then has the form $\vec A=(A_x,A_y,A_z)=\frac{B}{2}(y,-x,0)$.
With this vector potential,  Eqs. (\ref{eq:de31}) and
(\ref{eq:de32}) become
\begin{align}
(E-m v_f^2) \mid \psi_1 > &= v_f [(p_x -i\frac{\hbar
\tilde{\omega}^2}{v_f^2}x)- i(p_y - i\frac{\hbar
\tilde{\omega}^2}{v_f^2}y)] \mid \psi_2>, \label{eq:de33}\\
(E+m v_f^2) \mid \psi_1 > &= v_f [(p_x +i\frac{\hbar
\tilde{\omega}^2}{v_f^2}x)+ i(p_y + i\frac{\hbar
\tilde{\omega}^2}{v_f^2}y)] \mid \psi_2>,\label{eq:de34}
\end{align}
where $\tilde{\omega}^2=\omega_c^2/2$, with the cyclotron
frequency $\omega_c$ defined by $\omega_c^2=eBv_f^2/\hbar$. To
solve Eqs. (\ref{eq:de33}) and (\ref{eq:de34}), we again introduce
the right- and left-handed chiral annihilation and creation
operators \cite{AB07-2}. Then we write $r_j$ and $p_j$ in terms of
operators $\tilde{a}_j$ and $\tilde{a}_j^\dag$, from Eqs.
(\ref{eq:de33}) and (\ref{eq:de34}) we obtain
\begin{align}
\mid \psi_1 > &= -i \frac{2\hbar\tilde{\omega}}{E-m v_f^2}
\tilde{a}_r
\mid \psi_2>, \label{eq:de35}\\
\mid \psi_2 > &= i\frac{2\hbar\tilde{\omega}}{E+m v_f^2}
\tilde{a}_r^\dag \mid \psi_1>. \label{eq:de36}
\end{align}
In contrast to Eqs. (\ref{eq:de13}) and (\ref{eq:de14}), where
only the left-handed chiral operators appear, here in Eqs.
(\ref{eq:de35}) and (\ref{eq:de36}) only the right-handed chiral
operators are present\cite{AB07-2}. Substituting Eqs.
(\ref{eq:de35}) and (\ref{eq:de36}) into Eq. (\ref{eq:de3}), it is
possible to write the Hamiltonian of this system in the following
form
\begin{align}
H_2= - \hbar ( \tilde{g}_r \sigma^{+} \tilde{a}_r + \tilde{g}_r^*
\sigma^{-} \tilde{a}_r^\dag) +mv_f^2\sigma_z, \label{eq:Doham2}
\end{align}
with $\tilde{g}_r = 2i\tilde{\omega}=iv_f\sqrt{2eB/\hbar}$. The
Hamiltonian $H_2$ describes a massive electron of right-handed
chirality. From Eqs. (\ref{eq:de35}) and (\ref{eq:de36}), we
obtain the energy spectrum given by
\begin{align}
E_{n_r}= \pm \sqrt{2eB\hbar v_f^2\left(n_r + \frac{1}{2} \pm
\frac{1}{2}\right) + m^2v_f^4}, \label{eq:enes5}
\end{align}
with $n_r=0,1,2,\ldots$. For the case of massless charge carriers
the energy spectrum (\ref{eq:enes5}) is written as
\begin{align}
E_{n_r}= \pm \sqrt{2eB\hbar v_f^2\left(n_r + \frac{1}{2} \pm
\frac{1}{2}\right)}, \label{eq:enes6}
\end{align}
which corresponds to the well-known unusual Landau-level spectrum
of monolayer graphene \cite{GU2005,MC2006,PE2006}. We note that
the expression (\ref{eq:enes6}) has exactly the same form as the
Landau levels given by expression (1) of the reference
\cite{KA2007}. On the other hand, for the case $m\neq0$, the
non-relativistic limit of (\ref{eq:enes5}) leads to the following
result
\begin{align}
E^{n-r}_{n_r}= \pm \sqrt{2eB\hbar v_f^2\left(n_r + \frac{1}{2} \pm
\frac{1}{2}\right) + m^2v_f^4} - mv_f^2, \label{eq:enes7}
\end{align}
which corresponds to the Landau levels for bilayer graphene. This
Landau-level spectrum implies that one of the levels obtained for
$n_r=0$ has a $B$-dependent gap of the form
\begin{align}
\Delta E^{n-r}_{0}= -\delta + \sqrt{\gamma B + \delta^2},
\label{eq:enes7}
\end{align}
where $\delta=2mv_f^2$ and $\gamma = 8ev_f^2\hbar$. This result
has certain analogy with the experimental gap reported in
\cite{VE2012}. The non-relativistic limit of (\ref{eq:enes5})
implies that $\delta^2 \gg \gamma B$, thus the Landau levels can
be written as $E^{n-r}_{n_r}=eB/\hbar (n_r + \frac{1}{2} \pm
\frac{1}{2})$ and the $B$-dependent gap obtained for one of the
levels associated to $n_r=0$ is written as $\Delta
E^{n-r}_{0}=2eB\hbar/m$. We note that the charge carriers
described by Eq. (\ref{eq:de3}) have right-handed chirality.

For the full problem described by Eq. (\ref{eq:de2}), we consider
simultaneously the two systems described by Hamiltonians
(\ref{eq:Doham1}) and (\ref{eq:Doham2}). The Hamiltonian that
describes the dynamics of charge carriers of graphene in presence
of a uniform magnetic field is given by
\begin{align}
H = H_l - H_r + mv_f^2 \sigma_z, \label{eq:Doham3}
\end{align}
where the left-handed Hamiltonian $H_l$ is given by
\begin{align}
H_l = \hbar ( g_l \sigma^{+} a_l^\dag + g_l^* \sigma^{-} a_l ),
\label{eq:Doham4}
\end{align}
with $g_l = iv_f\sqrt{eB_I/\hbar}$, and the right-handed
Hamiltonian $H_r$ is given by
\begin{align}
H_r = \hbar ( \tilde{g}_r \sigma^{+} \tilde{a}_r + \tilde{g}_r^*
\sigma^{-} \tilde{a}_r^\dag), \label{eq:Doham5}
\end{align}
with $\tilde{g}_r = iv_f\sqrt{2eB/\hbar}$.

For the case of a weak external field $B \ll B_I$, in a first
approximation where corrections of order ${\it O}(B/B_I)$ are
neglected, Eq. (\ref{eq:de2}) leads to the energy spectrum given
by (\ref{eq:enes1}). On the other hand, for the case of a strong
external magnetic field $B \gg B_I$,  neglecting corrections of
order ${\it O}(B_I/B)$, the Eq. (\ref{eq:de2}) leads to the energy
spectrum of (\ref{eq:enes5}). This means that in the limit $B
\rightarrow 0$ the charge carriers have left-handed chirality,
while for the limit $B \rightarrow \infty$ they have right-handed
chirality. In other words, changing the strength of the external
magnetic field $B$ must lead to the existence of a chiral quantum
phase transition for the system described by Eq. (\ref{eq:de2}). A
complete characterization of this kind of chiral phase transition,
using the two-dimensional Dirac oscillator couplings in the
context of (Anti) Jaynes-Cummings models, is performed in
reference \cite{AB082} and the signatures of a possible quantum
phase transition for the case of a graphene quantum dot model in a
magnetic field has been reported in \cite{RO2012}.

Without considering the presence of the external magnetic field,
for the case of massless charge carriers (monolayer graphene), the
left-handed hamiltonian is given by (\ref{eq:Doham5}), while for
the case of massive charge carriers (bilayer graphene), the
left-handed hamiltonian is given by (\ref{eq:Doham1}). Both these
Hamiltonians can be described within a tight-binding model as
shown in \cite{FV2013}, for the one-dimensional Dirac oscillator.
This fact means that the Dirac oscillator description, that we
have introduced in this work, is consistent with the tight-binding
approximation that traditionally has been used to describe the
electronic structure of graphene \cite{GU2005,MC2006,PE2006}.

The results that we have presented in this letter have shown that
the two-dimensional Dirac oscillator describes consistently some
electronic properties of graphene. This is the first time that the
Dirac oscillator describes a very well know physics system. With
the present model, it has been possible to explain that the
left-handed chirality of charge carriers in monolayer and bilayer
graphene is originated by the linear potential that appears in the
Dirac oscillator model. This linear potential describes the
interactions of the medium over the charge carriers. We have
interpreted the parameter $B_I$ that appears in the linear
potential as an internal field acting over the charge carriers,
which is assumed to originate in an effective form from the motion
of the charge carriers relative to the planar hexagonal
arrangement of carbon atoms. This model leads consistently to the
very well known dispersion relations of charge carriers in
monolayer and bilayer graphene which traditionally are obtained
using a tight-binding approximation. For the case in which an
external and uniform magnetic field $B$ is presented in a
perpedicular form over the layers, the well known Landau level
spectrum for the monolayer graphene is obtained. For the case of
bilayer graphene, the model predicts the existence of a
$B$-dependent gap. Finally, this model also predicts that changing
the strength of $B$ must lead to the existence of a chiral quantum
phase transition for the system.

C. Quimbay thanks the School of Physical Sciences at the
University of Kent, Canterbury, for their hospitality during his
visit there.


\end{document}